\documentclass[a4paper]{article}
\usepackage{amsmath,amssymb,theorem,fullpage,mathtools}
\usepackage[shortcuts]{extdash}
\usepackage{hyperref}
\usepackage[nosort]{cite}

\let\frac\undefined

\allowdisplaybreaks

\numberwithin{equation}{section}

\def\Maketitle{{\def\newpage{}\maketitle}}

\def\eq#1$$#2$${\begin{equation#1}#2\end{equation#1}}
\long\def\subeq#1{\begin{subequations}#1\end{subequations}}
\def\Split$$#1$${\begin{split}#1\end{split}}
\def\Align#1$$#2$${\begin{align#1}#2\end{align#1}}
\def\AlignAt#1$$#2$${\begin{alignat}{#1}#2\end{alignat}}
\def\Aligned#1{\begin{aligned}#1\end{aligned}}
\def\Gather#1$$#2$${\begin{gather#1}#2\end{gather#1}}
\def\Gathered#1{\begin{gathered}#1\end{gathered}}
\def\Multline#1$$#2$${\begin{multline#1}#2\end{multline#1}}

\def\?{\notag}
\def\d{\partial}
\def\bd{\bar\partial}

\def\const{\mathop{\rm const}\nolimits}
\def\cA{{\cal A}}

\def\cO{{\cal O}}

\def\sh{\mathop{\rm sh}\nolimits}
\def\ch{\mathop{\rm ch}\nolimits}
\def\th{\mathop{\rm th}\nolimits}

\def\lcolon{\mathopen{\,:\,}}
\def\rcolon{\mathclose{\,:\,}}

\def\Z{{\mathbb Z}}

\def\e{{\rm e}}
\def\i{{\rm i}}

\def\ttau{{\tilde\tau}}

\def\bT{{\bar T}}
\def\bz{{\bar z}}

\def\bh{{\bar h}}

\makeatletter

\def\section{\@startsection{section}{1}{\z@}%
                                   {-3.5ex \@plus -1ex \@minus -.2ex}%
                                   {2.3ex \@plus.2ex}%
                                   {\normalfont\normalsize\bfseries}}
\def\subsection{\@startsection{subsection}{2}{\z@}%
                                     {-3.25ex\@plus -1ex \@minus -.2ex}%
                                     {1.5ex \@plus .2ex}%
                                     {\normalfont\normalsize\bfseries\itshape}}
\def\@seccntformat#1{\csname the#1\endcsname.~~}
\long\def\@makecaption#1#2{%
  \vskip\abovecaptionskip
  \sbox\@tempboxa{\small#1. #2}%
  \ifdim \wd\@tempboxa >0.9\hsize
  {\leftskip=0.05\hsize\rightskip=0.05\hsize\relax\small
    #1. #2\par}
  \else
    \global \@minipagefalse
    \hb@xt@\hsize{\hfil\box\@tempboxa\hfil}%
  \fi
  \vskip\belowcaptionskip}
\def\Appendix{\appendix
  \def\@seccntformat##1{Appendix~\csname the##1\endcsname.~~}}

\let\over\@@over
\let\atop\@@atop
\let\above\@@above
\let\overwithdelims\@@overwithdelims
\let\atopwithdelims\@@atopwithdelims
\let\abovewithdelims\@@abovewithdelims

\makeatother

\begin{document}

\title{Note on four-particle form factors of operators $T_{2n}T_{-2n}$ in sinh-Gordon model}
\author{Michael Lashkevich and Yaroslav Pugai\\[\medskipamount]
\parbox[t]{0.9\textwidth}{\normalsize\it\raggedright
Landau Institute for Theoretical Physics, 142432 Chernogolovka, Russia\medspace%
\footnote{Mailing address.}
\\
Kharkevich Institute for Information Transmission Problems, 19 Bolshoy Karetny per., 127994 Moscow, Russia\\
Moscow Institute of Physics and Technology, 141707 Dolgoprudny, Russia}
}
\date{}

\Maketitle

\begin{abstract}
The diagonal matrix elements $\langle\theta_1,\theta_2|T_{2n}T_{-2n}|\theta_1,\theta_2\rangle$ between two-particle states in the sinh-Gordon model are computed analytically for all integers $n>0$. This confirms the proposal~\cite{SZ-toappear:2014} by F.~Smirnov and A.~Zamolodchikov for these matrix elements and demonstrates effectiveness of the algebraic approach to form factors.
\end{abstract}

\section{Introduction}
We continue studying the structure of the descendant fields in the integrable two-dimensional quantum field theory in the algebraic approach \cite{Feigin:2008hs,Lashkevich:2013mca,Lashkevich:2013yja,Lashkevich:2014rua,Lashkevich:2014qna}. Our basic example is the sinh-Gordon model defined by the action
\eq$$
\cA_0=\int d^2x\,\left({(\d_\nu\varphi)^2\over16\pi}-2\mu\ch b\varphi\right).
\label{sh-G-action}
$$
The coupling constant $\mu$ has the scaling dimension $2+2b^2$ so that the perturbation is relevant. Integrability of the theory is related to the existence of an infinite set of  integrals of motion. The corresponding conservation laws for currents have the form~\cite{Zamolodchikov:1987jf}%
\footnote{We assume $z=x^1-x^0$, $\bz=x^1+x^0$ for the light-cone variables and $ds^2=(dx^0)^2-(dx^1)^2=-dz\,d\bz$ for the metrics in the Minkowski space. Correspondingly, $\d=\d/\d z$, $\bd=\d/\d\bz$.}
\eq$$
\Aligned{
\bd T_{2n}
&=\d\Theta_{2n-2},
\\
\d T_{-2n}
&=\bd\Theta_{-2n+2},
}\label{conservation-laws}
$$
for $n=1,2,\ldots\,$. As usual, the subscript at a current operator is its spin.

The spectrum of the model consists of a massive scalar boson with the following two-particle $S$ matrix
\eq$$
S(\theta)={\th{1\over2}(\theta-\i\pi r)\over\th{1\over 2}(\theta+\i\pi r)},
\label{S-matrix}
$$
where the parameter $r$ is related to the parameter $b$ as $b^2=(1-r)/r$. Correlation functions of local operators can be studied in the framework of the form factor approach. In this approach form factors, which are matrix elements of local operators with respect to eigenvectors of the Hamiltonian, are obtained as solutions to a system of bootstrap equations (see~\cite{Smirnov:1992vz} and references therein). Correlation function are expressed in terms of form factors by inserting a decomposition of unity in the sum over stationary states between local operators. In fact, this provides infrared (long distance) expansions of correlation functions.

In principle, all said is valid for the sine-Gordon model, which corresponds to $-1\le b^2<0$, but the spectrum of the latter is much richer. It consists of solitons and breathers, and the $S$-matrix (\ref{S-matrix}) only describes scattering of a pair of the lightest breather particles. Nevertheless, our derivation of the main formula (\ref{2-2-ff}) is only valid if the spectrum consists of one neutral scalar particle. This is the case for the sinh-Gordon theory and for the restricted sine-Gordon theory with $r={2s+1\over2s-1}$, $s=2,3,\ldots\,$, which corresponds to the $\Phi_{13}$ perturbations of the minimal conformal models with the central charge $c=-2{(6s-1)(s-1)\over2s+1}$~\cite{Smirnov:1990vm}.

Recently, F.~Smirnov and A.~Zamolodchikov proposed a class of new infinite-dimensional families of integrable effective (non-renormalizable) field theories \cite{SZ-toappear:2014}. These models can be defined as irrelevant perturbations of any integrable model by spinless products of conserved currents. In the case of the sinh- and sine-Gordon models the action of the perturbed models looks like:%
\footnote{Examples of such IR effective theories have already appeared in the literature. Al.~Zamolodchikov's RG flow from the tricrital Ising fixed point to the critical one is related with the two-dimensional Majorana massless fermion perturbed by the operator $T\bT=T_2T_{-2}$ \cite{Zamolodchikov:1987ti,Zamolodchikov:1991vx}. In the case of the massive fermion such kind of perturbations were studied by G.~Mussardo and P.~Simon~\cite{Mussardo:1999aj}.}
\eq$$
\cA=\cA_0-\sum_{n>0}\lambda_n
\int d^2x\,T_{2n}T_{-2n}(x).
\label{A-effective}
$$
Here the product $T_{2n}T_{-2n}(x)$ is understood as a limit~\cite{Zamolodchikov:2004ce}
$$
T_{2n}T_{-2n}(x)
=\lim_{\delta\to0}(T_{2n}(x+\delta)T_{-2n}(x)-\Theta_{2n-2}(x+\delta)\Theta_{-2n+2}(x)-\d_\mu J_{n,\text{sing}}^\mu(x,\delta)).
$$
The last term cancels the singular part, which is nothing but a total divergence. It means that the whole operator is defined modulo a total divergence, so that the perturbation terms in the action are well defined on the infinite plane.

Smirnov and Zamolodchikov found such models to be integrable and proposed an elegant expression for their exact $S$ matrices. An important part of their derivation of the exact $S$ matrix was the conjecture that the matrix element%
\footnote{Here, as usual, the state $|\theta_1\ldots\theta_N\rangle$ means the stationary state, which contains $N$ particles with the rapidities $\theta_1,\ldots,\theta_N$. The rapidity $\theta_i$ is related to the corresponding 2-momentum $p_i$ via $p_i^0=m\ch\theta_i$, $p_i^1=m\sh\theta_i$.}
$$
\langle\theta_1\theta_2|T_{2n}T_{-2n}(0)|\theta_3\theta_4\rangle
$$ 
in the limit
\eq$$
\theta_3\to\theta_1,
\qquad
\theta_4\to\theta_2
\label{specialpoint}
$$
after a proper regularization is given by
\eq$$
\langle\theta_1\theta_2|T_{2n}T_{-2n}(0)|\theta_1\theta_2\rangle
=4Z_n^2\sh\theta_{12}\sh(2n-1)\theta_{12},
\qquad
\theta_{12}=\theta_1-\theta_2.
\label{2-2-ff}
$$
The overall constant factor depends on the normalization of currents:
\eq$$
Z_n=\e^{-4n\theta}\langle\theta|T_{2n}(0)|\theta\rangle.
\label{Zn-def}
$$
For the simplest case $T\bar T\equiv T_2T_{-2}$ this formula can be checked by means of an explicit expression found by G.~Delfino and G.~Niccoli\cite{Delfino:2006te}. An expression for general values of $n$ was proposed in our recent paper~\cite{Lashkevich:2014qna}, where we proposed an algebraic construction for form factors of these operators. Though the corresponding explicit formula seems to be rather cumbersome, we show in this note that the algebraic construction provides the most direct way for an analytic proof of~(\ref{2-2-ff}). In the normalization, chosen in \cite{Lashkevich:2014qna}, the overall coefficient reads
\eq$$
Z_n=-{\pi m^{2|n|}\over2}{\cos{\pi r(2n-1)\over2}\over\cos{\pi r\over2}}.
\label{Zn-explicit}
$$
This normalization conforms with the natural normalization of the energy-momentum tensor component $T=-2\pi T_{zz}$ for $n=1$, but for general values of $n$ it is chosen arbitrarily. In addition, we show that no regularization is necessary for $n>1$, but the definition of $T\bT$ needs to be slightly modified.

\section{Calculation of $2-2$ matrix elements}

Our aim is to compute explicitly the four-point form factor $\langle\theta_1\theta_2|T_{2n}T_{-2n}(0)|\theta_1\theta_2\rangle$. Form factors of such kind of operators were found in~\cite{Lashkevich:2014qna} in the form of special matrix elements in an auxiliary Fock space. Let us recall some facts concerning form factors relevant to our aim. In what follows we use the notation of~\cite{Lashkevich:2014qna}.

Form factors of a local operator $\cO(x)$ are its matrix elements in the basis of eigenvectors of the Hamiltonian. On the infinite plane form factors can be expressed in terms of a set of analytic functions $F_\cO$ of rapidities:
\eq$$
\langle\theta'_1\ldots\theta'_K|\cO(0)|\theta_1\ldots\theta_L\rangle
=F_\cO(\theta_1,\ldots,\theta_K,\theta'_L-\i\pi,\ldots,\theta'_1-\i\pi),
\label{ff-def}
$$
if $\theta_1>\theta_2>\cdots>\theta_K$, $\theta'_1>\cdots>\theta'_L$. The functions $F_\cO(\theta_1,\ldots,\theta_N)$ for any $N$ can be found in the form
\eq$$
F_\cO(\theta_1,\ldots,\theta_N)=G_\cO\rho^NJ_\cO\left(\e^{\theta_1},\ldots,\e^{\theta_N}\right)
\prod^N_{i<j}R(\theta_i-\theta_j).
\label{ff-factorization}
$$
The function $R(\theta)$ and the constant $\rho$ are operator independent:
\eq$$
R(\theta)=\exp\left(
-4\int^\infty_0{dt\over t}\,
{\sh{\pi t\over2}\sh{\pi(1-r)t\over2}\sh{\pi rt\over2}
\over\sh^2\pi t}\ch(\pi+\i\theta)t
\right),
\qquad
\rho=\left(R(\i\pi)\sin\pi r\right)^{-1/2}.
\label{R-rho-def}
$$
The functions $J_\cO(x_1,\ldots,x_N)$ are symmetric and rational and define the operator $\cO(x)$. The normalization factor $G_\cO$ is separated for the sake of convenience. It accumulates some data that cannot be found in an algebraic way. As it was shown in~\cite{Feigin:2008hs} the $J_\cO$ can be expressed as linear combinations of matrix elements of the form
\eq$$
J^{h\bh'}_a(x_1,\ldots,x_N)
={}_a\langle h|t(x_1)\cdots t(x_N)|h'\rangle_a,
\label{Jhh'-def}
$$
where the stated ${}_a\langle h|$ and $|h'\rangle_a$ belong to an auxiliary Fock spaces, while the current $t(z)$ is acting on these spaces and can be expressed in terms of the corresponding Heisenberg algebra generators. To give a precise meaning to this expression, define the Heisenberg algebra. Its generators are $\d^\pm_k$ ($k\in\Z\setminus\{0\}$), $\hat a$, $\d_a$ with the only nonvanishing commutation relations
\eq$$
[\d_a,\hat a]=1,
\qquad
[d^+_k,d^-_{-k}]=kA_k,
\qquad
A_k=(q^{k/2}-q^{-k/2})(q^{k/2}-(-)^kq^{-k/2}),
\label{dda-commut}
$$
where $q=\e^{-\i\pi r}$. Define the vacuums
\eq$$
{}_a\langle1|d^\pm_{-k}=0,
\quad
d^\pm_k|1\rangle_a=0
\quad(k>0),
\qquad
\hat a|1\rangle_a=a|1\rangle_a.
\label{vac-def}
$$
The corresponding normal ordering operation $\lcolon\cdot\rcolon$ puts $d^\pm_k$ ($k>0$) to the right of $d^\pm_{-k}$. Let
\eq$$
\lambda_\pm(z)=\exp\sum_{k\ne0}{d^\pm_k\over k}z^{-k}.
\label{lambdapm-def}
$$
The products of these operators are reduced to normal products according to
\eq$$
\lambda_{\varepsilon'}(z')\lambda_\varepsilon(z)
=\langle\lambda_{\varepsilon'}(z')\lambda_\varepsilon(z)\rangle
\lcolon\lambda_{\varepsilon'}(z')\lambda_\varepsilon(z)\rcolon,
\label{lambdalambda-prod}
$$
where
\eq$$
\langle\lambda_{\varepsilon'}(z')\lambda_\varepsilon(z)\rangle
=1+(q-q^{-1}){\varepsilon-\varepsilon'\over2}{z'z\over z^{\prime\,2}-z^2}.
\label{lambdalambda-pair}
$$
The current $t(z)$ is defined as (we assume $\varepsilon=\pm1$ and $\pm$ be the same)
\eq$$
t(z)=\sum_{\varepsilon=\pm1}\e^{-\i\pi\varepsilon\hat a}\lambda_\varepsilon(z).
\label{t(z)-def}
$$
Let $\cA$ be the commutative algebra with the generators $c_{-k}$, $k=1,2,\ldots\,$. We need two representations of this algebra
\eq$$
\pi(c_{-k})={d^-_k-d^+_k\over A_k},
\qquad
\bar\pi(c_{-k})={d^-_{-k}-d^+_{-k}\over A_k}.
\label{pick-def}
$$
For any element $h\in\cA$ we define the following vectors
\eq$$
{}_a\langle h|={}_a\langle1|\pi(h),
\qquad
|h\rangle_a=\bar\pi(h)|1\rangle_a.
\label{hvec-def}
$$
This completes the definition of the functions $J^{h\bh'}_a(x_1,\ldots,x_N)$.

The functions are easily calculated in the following way. First, decompose the expression (\ref{Jhh'-def}) by means of~(\ref{t(z)-def}):
$$
J^{h\bh'}_a(x_1,\ldots,x_N)
=\sum_{\varepsilon_1,\ldots,\varepsilon_N}\e^{-\i\pi a\sum_i\varepsilon_i}
\>{}_a\langle h|\lambda_{\varepsilon_1}(x_1)\cdots\lambda_{\varepsilon_N}(x_N)|h'\rangle_a.
$$
Second, use (\ref{hvec-def}) and push $\pi(h)$ to the right and $\bar\pi(h')$ to the left by means of the commutation relations
\subeq{\label{pibpi-props}
\Align$$
[\pi(c_{-k}),\lambda_\varepsilon(z)]
&=(-\varepsilon)^{k+1}z^k\lambda_\varepsilon(z),
\label{pilambda-commut}
\\
[\bar\pi(c_{-k}),\lambda_\varepsilon(z)]
&=-\varepsilon^{k+1}z^{-k}\lambda_\varepsilon(z),
\label{bpilambda-commut}
\\
[\pi(c_{-k}),\bar\pi(c_{-l})]
&=-(1+(-1)^k)kA^{-1}_k\delta_{kl},
\label{pibpi-commut}
$$}
Only the commutators contribute the result, since the remainder vanishes due to the relations
\eq$$
\pi(c_{-k})|1\rangle_a=0,
\qquad
{}_a\langle1|\bar\pi(c_{-k})=0.
\label{pibpi-zero}
$$
The result reduces to a combination (with $x_i$-dependent coefficients) of the matrix elements
$$
{}_a\langle1|\lambda_{\varepsilon_1}(x_1)\cdots\lambda_{\varepsilon_N}(x_N)|1\rangle_a
=\prod_{i<j}\langle\lambda_{\varepsilon_i}(x_i)\lambda_{\varepsilon_j}(x_i)\rangle,
$$
which are known due to~(\ref{lambdalambda-pair}).

We now return to the physics of the problem. The matrix elements $J_a(x_1,\ldots,x_N)\equiv J^{1\bar1}_a(x_1,\ldots,x_N)$ provide~\cite{Fring:1992pj,Koubek:1993ke,Lukyanov:1997bp} (by means of (\ref{ff-factorization})) form factors of the exponential operator
$$
V_a(x)=\e^{(b+b^{-1})(1/2-a)\varphi(x)},
$$
if the corresponding constant $G_\cO$ is its vacuum expectation value~\cite{Lukyanov:1996jj}. The functions $J^{h\bh'}_a(x_1,\ldots,x_N)$ were argued~\cite{Feigin:2008hs} to describe descendants of the operator~$V_a(x)$. We are interested in the operators $\cO_n=T_{2n}T_{-2n}$, which are special $(2n,2n)$-level descendants of the operator $V_{1/2}(x)$. In~\cite{Lashkevich:2014qna} we proposed that their form factors have the form (\ref{ff-factorization}) with the normalization
\eq$$
G^{T\bT}_n={\pi^2m^{4n}\over64\sin^2\pi r}
\label{Gn-def}
$$
and and the $J$-functions
\eq$$
J^{T\bT}_n(x_1,\ldots,x_N)={}_{3/2-r}\langle1|S^+_{2n}t(x_1)\ldots t(x_N)S^+_{-2n}|1\rangle_{r-1/2}+\cdots.
\label{JTT-def}
$$
The normalization is chosen in consistency with the normalization of the operators $T_{2n}(x)$ given by~(\ref{Zn-explicit}). It is only fixed by a physical condition for $n=1$. The dots in the r.h.s.\ of (\ref{JTT-def}) mean the terms that contain the factors of the form $\sum x_i^{2k-1}=\sum\e^{(2k-1)\theta_i}$ for some values of $k$, which vanish in the matrix elements
$$
\langle\theta_1\theta_2|T_{2n}T_{-2n}(0)|\theta_1\theta_2\rangle
=F_n(\theta_1,\theta_2,\theta_2-\i\pi,\theta_1-\i\pi).
$$
In other words, these irrelevant terms correspond to commutators of integrals of motion with some local operators.

The operators $S^+_k$ are so called inverse screening operators defined by
\eq$$
S^+_k=\oint{dz\over2\pi\i}\,z^{k-1}S^+(z),
\qquad
S^+(z)=\e^{(r-1)\d_a}\lcolon\exp\sum_{k\ne0}{d^+_k-d^-_k\over k(q^{k/2}-q^{-k/2})}z^{-k}\rcolon.
\label{S+-def}
$$
Since these operators are expressed in terms of $d^-_l-d^+_l$, they generate vectors of the form~(\ref{hvec-def}).

The main technical difficulty of calculating the matrix elements $J_n$ is that the form of the vectors ${}_{3/2-r}\langle1|S^+_{2n}$ and $S^+_{-2n}|1\rangle_{r-1/2}$ is rather cumbersome. They can be expressed in terms of Hall\--Littlewood polynomials (similar to null-vectors in~\cite{Lashkevich:2013yja}), but this does not simplify calculation of the matrix element. Instead, we push $S^+_{2n}$ to the right and $S^+_{-2n}$ to the left by commuting them with the currents~$t(x_i)$ and with each other.

To do it we need some properties of the operators~$S^+_k$. Namely, we use the commutation relations
\eq$$
S^+_kS^+_l=-S^+_{l+2}S^+_{k-2}
\label{S+S+-commut}
$$
and
\eq$$
[S^+_k,t(z)]
=(-1)^kB_1z^k\left(q^{-{k-1\over2}}\lcolon\eta(-q^{-1/2}z)\ttau(z)\rcolon\e^{\i\pi\hat a}
+q^{k-1\over2}\lcolon\eta(-q^{1/2}z)\ttau(z)\rcolon\e^{-\i\pi\hat a}\right).
\label{S+t-commut}
$$
Here $B_1=q^{1/2}+q^{-1/2}$ and
\Align$$
\ttau(z)
&=\e^{(r-1)\d_a}\lcolon\exp\sum_{k\ne0}{q^{k/2}d^+_k-q^{-k/2}d^-_k\over k(q^{k/2}-q^{-k/2})}z^{-k}\rcolon,
\label{tildetau-def}
\\
\eta(z)
&=\exp\sum_{k\in2\Z+1}{2(d^+_k-d^-_k)\over k(q^{k/2}-q^{-k/2})}z^{-k}.
\label{eta-def}
$$
The operator $\ttau(z)$ is quite analogous to the operator $\tau(z)$ introduced in (\cite{Lashkevich:2014qna}), and can be obtained from the latter by the substitution $r\to1-r$.

Besides, the commutations of $S^+_k$ with $\eta(z)$ and $\ttau(z)$ are trivial
\eq$$
S^+_k\eta(z)=\eta(z)S^+_k,
\qquad
S^+_k\ttau(z)=z^2\ttau(z)S^+_{k-2}.
\label{S+eta-ttau-commut}
$$
We also need the properties
\eq$$
\Gathered{
S^+_k|1\rangle_a=0,
\qquad
{}_a\langle1|S^+_{-k}
\quad(k>0),
\\
S^+_0|1\rangle_a=|1\rangle_a,
\qquad
{}_a\langle1|S^+_0={}_a\langle1|.
}\label{S+vac-prop}
$$
The operator $\eta(z)$ is such that it only produce a factor in the matrix element. Indeed, its pair correlation functions with $\lambda_-(z)$, $\lambda_+(z)$ and $\ttau(z)$ are the same:
\eq$$
\langle\eta(z')\lambda_\pm(z)\rangle=\langle\eta(z')\ttau(z)\rangle
=f_\eta\left(z\over z'\right),
\label{eta-corr}
$$
where
\eq$$
f_\eta(z)={(1-q^{1/2}z)(1-q^{-1/2}z)\over(1+q^{1/2}z)(1+q^{-1/2}z)}.
\label{feta-def}
$$
After substituting the commutation relations (\ref{S+S+-commut}), (\ref{S+t-commut}), (\ref{S+eta-ttau-commut}) and the property (\ref{S+vac-prop}) into (\ref{JTT-def}) and applying (\ref{eta-corr}) we obtain
\Align$$
J^{T\bT}_n(x_1,\ldots,x_N)
&=\sum^N_{i\ne j}F_n(x_i,x_j|\hat X_{i,j})
{\>}_{3/2-r}\langle1|\ttau(x_i)t(\hat X_{i,j})\ttau(x_j)|1\rangle_{r-1/2}
\notag
\\
&\quad
+\delta_{n1}\left(\delta_{N0}-{}_{r-1/2}\langle1|t(X)|1\rangle_{r-1/2}-{}_{3/2-r}\langle1|t(X)|1\rangle_{3/2-r}\right)+\cdots,
\label{JTT-tildetau}
$$
Here $\hat X_{i,\ldots}=X\setminus\{x_i,\ldots\}$ and
\eq$$
F_\sigma(x,y|X)
=-B_1^2x^{\sigma-1}y^{1-\sigma}G_\sigma(x|y,X)G_{-\sigma}(y|x,X),
\label{Fsigma-def}
$$
where
\eq$$
G_\sigma(x|X)
=q^{\sigma/2}\prod^N_{i=1}f_\eta\left(-q^{1/2}{x_i\over x}\right)
-q^{-\sigma/2}\prod^N_{i=1}f_\eta\left(-q^{-1/2}{x_i\over x}\right).
\label{Gsigma-def}
$$

The matrix element in the first line of~(\ref{JTT-tildetau}) is given by
\Multline$$
{}_{3/2-r}\langle1|\ttau(y_1)t(x_1)\ldots t(x_{N-2})\ttau(y_2)|1\rangle_{r-1/2}
\\
=\i^{N-2}\langle\ttau(y_1)\ttau(y_2)\rangle\sum_{\varepsilon_1,\ldots,\varepsilon_{N-2}}
\prod_i(-1)^{\varepsilon_i}\langle\ttau(y_1)\lambda_{\varepsilon_i}(x_i)\rangle\,
\langle\lambda_{\varepsilon_j}(x_i)\ttau(y_2)\rangle
\prod_{i<j}\langle\lambda_{\varepsilon_i}(x_i)\lambda_{\varepsilon_j}(x_j)\rangle,
\label{tautau-matel}
$$
where
\eq$$
\langle\ttau(z')\ttau(z)\rangle
={(z'-qz)(z'-z)(z'-q^{-1}z)\over z^{\prime2}(z'+z)},
\qquad
\langle\ttau(z')\lambda_\varepsilon(z)\rangle
=\langle\lambda_{-\varepsilon}(z')\ttau(z)\rangle
={z'-q^{-\varepsilon}z\over z'+z}.
\label{taulambda-matel}
$$
After substituting it into (\ref{JTT-tildetau}) for $N=4$ and simplifying (with the help of \emph{Mathematica$\,^\circledR$}) we make sure that the first line in the expression (\ref{JTT-tildetau}) for the function $J^{T\bT}_\sigma(x_1,\ldots,x_4)$ is regular at the point~(\ref{specialpoint}). The only singularity can come from the last line, which vanishes for $n\ge2$.

In the case $n\ge2$, after multiplying the answer by
\eq$$
\rho^4R(\theta_{12})R(-\theta_{12})R(\theta_{12}+\i\pi)R(-\theta_{12}+\i\pi)R^2(\i\pi)
={1\over\sin^2\pi r}{\sh^2\theta_{12}\over\sh^2\theta_{12}+\sin^2\pi r}
\label{rhoR-simplification}
$$
and by $G^{T\bT}_n$, we obtain~({\ref{2-2-ff}) with the normalization factor~(\ref{Zn-explicit}).

The case $n=1$ must be considered separately. Denote $(T\bT)_{\rm red}$ the operator defined by the first line of~(\ref{JTT-tildetau}). This operator is the operator $T\bT$ plus $\const\times\left(\Theta-{1\over2}\langle\Theta\rangle\right)$, which is proportional to the operator defined by the terms at the Kronecker symbol~$\delta_{n1}$. Since, by definition $\langle(T\bT)_{\rm red}\rangle=0$, we easily establish the coefficient:
\eq$$
(T\bT)_{\rm red}=T\bT+2\langle\Theta\rangle\Theta-\langle\Theta\rangle^2.
\label{TbTred-def}
$$
Hence, the definition (\ref{A-effective}) of the effective field theory should be modified
\eq$$
\cA=\cA_0-\lambda_1\int d^2x\,(T\bT)_{\rm red}-\sum_{n\ge2}\lambda_m\int d^2x\,T_{2n}T_{-2n}.
\label{A-effective-modified}
$$

\section{Conclusion}

In the previous papers \cite{Lashkevich:2013yja,Lashkevich:2014rua,Lashkevich:2014qna} we have shown that the algebraic approach to form factors makes it possible to derive some general identities for form factors independently of the number of particles, i.e.\ identities for the corresponding operators. In this note we show that it allows one to obtain a general result of another kind: we consider a very special matrix element, but obtain a formula for arbitrary spin $2n$ of the current. Both kinds of formulas demonstrate the main advantage of this approach: representing an explicit but complicated formula as a matrix element of rather homogeneous objects like currents with simple properties allows one to guess and prove identities in their general form, which seems absolutely impossible, if you look at the corresponding explicit expressions.

\section*{Acknowledgments}

We are grateful to F.~Smirnov and A.~Zamolodchikov for sharing with us their unpublished results, and their kind permission to publish this note before their paper is ready. The work was performed under the grant \#14--15--00150 of the Russian Science Foundation.

\raggedright

\end{document}